\documentclass[twocolumn,prb]{revtex4}
\usepackage{amsfonts,amsmath,amssymb,mathrsfs}
\usepackage[dvips]{graphicx,color}
\input epsf

\newcommand{\dir}{Figs}
\newcommand{\etal}{{\em et al.} }
\newcommand{\eg}{{\em e.~g.}, }
\newcommand{\ie}{{\em i.~e.}, }

\newcommand{\Tr}{\mathop{\rm Tr}\nolimits}

\begin{document}
\title{Polymer adsorption onto random planar surfaces: \\
Interplay of polymer and surface correlations}

\author{Alexey Polotsky}
\email{polotsky@Physik.Uni-Bielefeld.de} \affiliation{Fakult\"at
f\"ur Physik, Universit\"at Bielefeld, Universit\"atsstra{\ss}e
25, D-33615 Bielefeld, Germany}

\author{Friederike Schmid}
\affiliation{Fakult\"at f\"ur Physik, Universit\"at Bielefeld,
Universit\"atsstra{\ss}e 25, D-33615 Bielefeld, Germany}

\author{Andreas Degenhard}
\affiliation{Fakult\"at f\"ur Physik, Universit\"at Bielefeld,
Universit\"atsstra{\ss}e 25, D-33615 Bielefeld, Germany}

\begin{abstract}
We study the adsorption of homogeneous or heterogeneous polymers
onto heterogeneous planar surfaces with exponentially decaying
site-site correlations, using a variational reference system
approach. As a main result, we derive simple equations for the
adsorption-desorption transition line. We show that the adsorption
threshold is the same for systems with quenched and annealed
disorder. The results are discussed with respect to their
implications for the physics of molecular recognition.
\end{abstract}

\maketitle

\section{Introduction}
\label{Introduction}

Polymer adsorption has been studied intensively in the last
decades because of its wide practical applications and rich
physics~\cite{Eisenriegler,Fleer}. The early studies mostly
focused on the adsorption onto physically and chemically
homogeneous surfaces. In nature, however, real objects are very
rarely perfectly homogeneous with ideal geometry. Real substrates
usually contain some heterogeneity. Therefore, a number of more
recent studies have considered the influence of random
heterogeneities on the adsorption transition~\cite{
Balazs/Huang:1991, Balazs/Gempe:1991, Baumgaertner:1991,
Sebastian:1993, Sumithra:1994, Andelman:1993, Linden:1994,
Huber:1998, Srebnik:1996, Bratko:1997, Bratko:1998,
Chakraborty:1998, Srebnik:1998, Golumbfskie:1999, Srebnik:2000,
Srebnik:2001, Chakraborty_rev, Moghaddam:2002, Lee:2003,
Moghaddam:2003, Genzer:PRE:2001, Genzer:ACIS:2001,
Genzer:JCP:2001, Genzer:MTS:2002}.

From a more fundamental point of view, the adsorption of random
heteropolymers (RHPs) onto heterogeneous surfaces has attracted
much attention, because RHPs serve as biomimetic polymers to study
the physics of proteins and nucleic acids. Studying the
interaction of RHPs with random heterogeneous surfaces can provide
insight into the physics of molecular recognition, a process
playing a key role in living organisms (\eg enzyme-substrate or
antigen-antibody interactions).

The present work deals with the adsorption of homopolymers and
heteropolymers on surfaces with chemical heterogeneities. As a
general rule, the introduction of chemical heterogeneities at
constant average load promotes adsorption, since the polymers can
adjust their conformations such that they touch predominantly the
most attractive surface sites. The adsorption transition is thus
shifted to a lower value of the mean polymer/surface affinity. The
extent of the shift is determined by the characteristics of the
randomness. In the present work, we focus on the influence of
correlations on the adsorption transition.

Our study was motivated by the idea that clustering and the interplay of cluster sizes should
contribute to molecular recognition. This idea is supported, among other, by a series of recent
papers by Chakraborty and coworkers~\cite{ Srebnik:1996, Bratko:1997, Bratko:1998, Chakraborty:1998,
Srebnik:1998, Golumbfskie:1999, Srebnik:2000, Srebnik:2001, Chakraborty_rev}.
Using different analytical approaches and Monte Carlo (MC) simulations, these authors have
established that below the adsorption threshold, statistically blocky polymers favor
statistically blocky surfaces over uncorrelated surfaces.
They found that, upon increasing the strength of the interactions, the adsorption transition
of heteropolymers on heterogeneous surfaces is followed by a second sharp transition, where
the polymers freeze into conformations that match the surface pattern.
This happens at adsorption energies significantly beyond the adsorption transition.
In the biological context, a two-step adsorption process where adsorption is followed by
``freezing'' possibly describes the physics of protein/DNA recognition, where the protein
slides on the DNA for a while before finding it's specific docking site.
For other molecular recognition processes, such as antibody-antigen interactions, it is not
only important that a protein adsorbs to a given particle, but equally crucial that it does
{\em not} adsorb to other objects. In order to assess the importance of cluster size matching
in those cases, one must calculate the shift of the adsorption transition
as a function of correlation lengths on the polymer and on the surface.

In a previous publication~\cite{Polotsky:2003}, we have calculated
the adsorption/desorption transition of an ideal
(non-self-interacting) RHP with sequence correlations on an
impenetrable planar homogeneous surface. We have compared two
different approaches. First, we have calculated numerically the
partition function of a random sample of lattice
RHPs~\cite{Fleer}. Provided the sample is sufficiently large, this
approach is basically exact. Second, we have applied a reference
system approach which was originally developed by
Chen~\cite{Chen:2000} and Denesyuk and
Erukhimovich~\cite{Denesyuk:2000} in a similar context (RHP
localization at liquid-liquid interfaces). The second approach
allowed to derive an explicit and surprisingly simple expression
for the adsorption transition point (repeated here in 
Eq.~(\ref{trp_rhp_final})),
which fitted the numerical results excellently.

Having thus established the validity of the reference system approach, we use
it here to study the adsorption of a single Gaussian chain onto a heterogeneous
planar surface with exponentially decaying site-site correlations. We consider
both homopolymers and random heteropolymers.
As in the case of the RHP on an homogeneous surface, we shall derive simple
analytic formulae for the adsorption/desorption transition point. This will
allow to discuss how surface patchyness influences the adsorption behavior,
and whether the interplay of surface and polymer correlations can bring about
something relevant to molecular recognition in the vicinity of the adsorption transition.

The problem of adsorption on heterogeneous surfaces has already been considered
theoretically by a number of authors, both for homopolymers and heteropolymers
(block-copolymers, periodic or random copolymers).

Baumg\"artner and Muthukumar~\cite{Baumgaertner:1991} studied the effect
of chemical and physical surface roughness on adsorbed homopolymers. Using MC techniques,
they calculated the adsorbed homopolymer amount on chemically heterogeneous surfaces,
as a function of temperature and determined the shift of the adsorption-desorption
transition temperature with decreasing concentration of adsorbing sites for different
chain lengths. In addition, they presented a scaling analysis showing that lateral size of
the adsorbed chain $R_{||}$ depends on the distribution of repelling "impurities" on the
adsorbing surface and the strength of their repulsive interaction with the polymer.
$R_{||}$ is unaffected by randomness and simply equal to the radius of gyration of a Gaussian
chain in two dimensions if impurities are very rare or weak. However, when the impurities are
not rare and repulsive barriers are high, $R_{||}$ is of order of the mean distance between
neighboring impurities.

Similar conclusions were reached by Sebastian and
Sumithra~\cite{Sebastian:1993, Sumithra:1994}, who developed an
ana\-ly\-ti\-cal theory of the adsorption of Gaussian chains on
random surfaces, using path integral methods combined with the
replica trick and a Gaussian variational approach. They took
surface heterogeneity into account by modifying de Gennes'
adsorption boundary condition~\cite{deGennes:1978, deGennes_book},
and analyzed the influence of randomness on the conformation of
the adsorbed chains.

MC simulations of single polymer adsorption have also been performed by Moghaddam and
Wittington~\cite{Moghaddam:2002}. The sites on the heterogeneous surface were taken to
be sticky or neutral for the polymer. The temperature dependencies of the energy and
the heat capacity were calculated for different sticker fractions and compared for the
cases of quenched and annealed surface disorder. Huber and Vilgis~\cite{Huber:1998}
have studied the case of physically rough surfaces.

The above mentioned works dealt with the single-chain problem. A number of other authors
investigated the adsorption from solution onto heterogeneous surfaces using MC
simulations~\cite{Balazs/Huang:1991, Balazs/Gempe:1991}, numerical lattice
self-consistent field (SCF) calculations~\cite{Linden:1994}, and an analytical SCF
theory~\cite{Andelman:1993}.

As mentioned earlier, the problem of heteropolymer adsorption on
heterogeneous surfaces has been studied extensively by
A.~Chakraborty and coworkers. Most of the results were summarized
in a recent review article~\onlinecite{Chakraborty_rev}. Srebnik
\etal used replica mean field calculations and the replica
symmetry breaking scheme to study the adsorption of Gaussian RHP
chains for \emph{theta} solvents~\cite{Srebnik:1996} and for good
solvents~\cite{Srebnik:2001}. MC simulations combined with simple
non-replica calculations allowed to analyze the role of the
statistics characterizing the sequence and surface site
distribution, and that of intersegment interactions, for the
pattern recognition between a disordered RHP and a disordered
surface~\cite{Bratko:1997, Bratko:1998, Chakraborty:1998,
Srebnik:1998, Golumbfskie:1999, Srebnik:2000}.

The adsorption of globular RHPs (\eg RHPs in a bad solvent) on patterned surfaces has
recently been considered by Lee and Vilgis with a variational approach~\cite{Lee:2003}.
A MC simulation study of single RHP chain adsorption onto random surfaces has also been
performed by Moghaddam~\cite{Moghaddam:2003}.

Interesting applications of interactions between copolymers and heterogeneous surfaces
have been examined by Genzer \cite{Genzer:PRE:2001, Genzer:ACIS:2001, Genzer:JCP:2001,
Genzer:MTS:2002}. Using a three dimensional self-consistent field model, he studied
the adsorption of A-B copolymers from a mixture of A-B copolymers and
A homopolymers onto a heterogeneous planar substrate composed of two chemically distinct
regularly or randomly distributed sites. He demonstrated that via substrate recognition,
the copolymer can transcribe the two-dimensional surface motif into three dimensions.
The specific way the surface motif is translated was shown to be strongly dictated by
the copolymer sequence.

The present work differs from all these approaches in that we focus on the
adsorption transition, \ie on the region in parameter space where
adsorption just sets in, and that we derive simple analytical expressions
for the location of that transition. We shall see that the characteristic
features observed in the regime of stronger adsorption do not necessarily
carry over to this regime of weak adsorption.

The rest of the paper is organized as follows: Sec.~\ref{Model} introduces the model.
The free energy calculation based on the reference system approach is explained and
carried out in Sec.~\ref{reference}. Some aspects of the calculations in this section
are very similar to those already described in our earlier publication,
Ref.~\onlinecite{Polotsky:2003}. Therefore, we will skip the corresponding intermediate
technical steps and refer the reader to Ref.~\onlinecite{Polotsky:2003} for more details.
Explicit analytic expressions for the adsorption transition point are obtained in
Sec.~\ref{Transition}. We summarize and discuss our final expressions in Sec.~\ref{Conclusions}.
Some intermediate calculations are described in three Appendices.

\section{The Model}
\label{Model}

We consider single Gaussian polymers chain consisting of $N$ monomer units near an
impenetrable planar surface. The surface carries randomly distributed sites of different
chemical nature. The surface pattern is described by a locally varying interaction
parameter $\sigma({\bf x})$. Here ${\bf x}$ denotes Cartesian coordinates $(x,y)$.
A particular (hetero)polymer realization is described by a sequence $\xi(n)$,
where $n$ indicates the monomer number ($n \in [0:N]$). The strength and the sign
of the monomer-surface interaction is determined by the product $\xi(n) \: \sigma({\bf x})$
(a "+" sign refers to attraction, ''--'' to repulsion).

The Hamiltonian of the system can, therefore, be represented as follows:
\begin{equation}
\label{Hamiltonian}
\frac{H}{kT}=\frac{3}{2a^2}\int_0^N
dn\left(\frac{\partial {\bf r}}{\partial n}\right)^2-\beta\int_0^N
dn \: V(z(n)) \: \xi(n) \: \sigma({\bf x}(n)) \, ,
\end{equation}
where ${\bf r} (n)$ is the chain trajectory in the space, ${\bf r}
\equiv \{{\bf x}, z\}$, $\beta=1/kT$ is the Boltzmann factor, $a$ is the Kuhn segment
length of the chain, and $V(z)$ describes the shape of the monomer-surface interaction
potential. This potential is taken to be a delta-function pseudopotential shifted at a
small but finite distance $z_0$ from the impenetrable surface and has the form
\begin{equation}
\label{potential}
V(z)=\left\{
\begin{array}{lc}
\delta(z-z_0) & \mbox{if } z>0 \\
+ \infty &  \mbox{if } z<0
\end{array}
\right.  \, .
\end{equation}

The model Hamiltonian (\ref{Hamiltonian}) can be used to describe
interactions of both heteropolymers and homopolymers with both
heterogeneous and homogeneous surfaces. Homopolymers are described
by a constant sequence function, i.e. $\xi(n) \equiv \xi_0$. The
sequences $\xi(n)$ of heteropolymers are taken to be randomly
distributed according to a Gaussian distribution function
\begin{eqnarray}
\label{seq_prob}
\lefteqn{
P\{\xi(n)\} \propto
} \\ &&
\exp \! \left[ -\frac{1}{2} \!\int \!\!\!\!\! \int_0^N 
\! \!\! dn_1 \: dn_2 (\xi(n_1)\!-\!\xi_0)
\: c^{-1}(n_1,n_2) \: (\xi(n_2)\!-\!\xi_0) \right],
\nonumber
\end{eqnarray}
where $\xi_0 = \langle \xi \rangle$ characterizes the ``mean charge'' of the chain,
$c(n_1,n_2)$ describes sequence correlations, and $c^{-1}$ is the inverse of $c$,
defined by $\int \! dn \, c^{-1}(n_1,n)c(n,n_2)=\delta(n_1-n_2)$.
Here we consider exponentially decaying correlations, {\em i.~e.},
\begin{eqnarray}
\label{seq_corrf} c(n_1, n_2)&\equiv&
\left\langle(\xi(n_1)-\xi_0)(\xi(n_2)-\xi_0)\right\rangle
\\
&=&c(|n_2-n_1|)=\Delta_p^2
\: e^{-\Gamma_p |n_2-n_1|} \, .
\nonumber
\end{eqnarray}
The parameter $\Gamma_p$ corresponds to an inverse ``contour correlation length``
on the polymer, and $\Delta_p$ gives the variance of the single-monomer distribution.
The proportionality factor in Eq.~(\ref{seq_prob}) is chosen such that $P\{\xi(n)\}$
is normalized. This heteropolymer model also covers homopolymers in the limit $\Delta_p \to 0$.

Similarly, the distribution of surface sites on heterogeneous surfaces is chosen
to be Gaussian, with the probability
\begin{eqnarray}
\label{surf_prob} 
\lefteqn{
P\{\sigma({\bf x})\} \propto 
}
\\&&
\exp \! \left[ -\frac{1}{2} \! \int \!\!\!\!\!\! \int \!\! 
d{\bf x}_1 \: d{\bf x}_2 \:
(\sigma({\bf x}_1)\!-\!\sigma_0)
\: g^{-1}({\bf x}_1, {\bf x}_2)\:
(\sigma({\bf x}_2)\!-\!\sigma_0)\right] \, ,
\nonumber
\end{eqnarray}
where $\sigma_0=\langle\sigma \rangle$ is the mean load of the surface, and the
site-site correlation function $g({\bf x}_1, {\bf x}_2)$ decays exponentially:
\begin{eqnarray}
\label{surf_corrf} g({\bf x}_1, {\bf x}_2)&\equiv&
\left\langle(\sigma({\bf x}_1)-\sigma_0)(\sigma({\bf
x}_2)-\sigma_0) \right\rangle
\\ &=&
g(|{\bf x}_1-{\bf x}_2|)= \Delta_s^2
e^{-\Gamma_s |{\bf x}_1-{\bf x}_2|} \, .
\nonumber
\end{eqnarray}
Here $\Delta_s$ is again the variance of the single-site
distribution on the surface and $\Gamma_s$ is the inverse
correlation length of the surface pattern. As before, $g^{-1}({\bf
x}_1,{\bf x}_2)$ is the inverse function of $g({\bf x}_1,{\bf
x}_2)$ with the property $\int d{\bf x} \, g^{-1}({\bf x}_1,{\bf
x})g({\bf x},{\bf x}_2)=\delta({\bf x}_1-{\bf x}_2)$, and
$P\{\sigma({\bf x})\}$ is normalized. Homogeneous surfaces are
obtained in the limit $\Delta_s \to 0$, and are characterized by
$\sigma({\bf x}) \equiv \sigma_0$.

\section{The Reference System Approach}
\label{reference}

\subsection{Quenched Average via Replica Trick}

The free energy of the system is given by the logarithm of the
conformational statistical sum averaged over all realizations of
the disorder $ \beta F=\langle \ln Z \rangle_{\{ \sigma({\bf
x}),\xi(n) \}}$. Here and below angular brackets $\langle \ldots
\rangle$ stand for disorder average. To perform the quenched
average we use the well-known replica trick
\begin{equation}
\label{replica}
\langle \ln Z \rangle_{\{ \sigma({\bf x}), \xi(n) \} } = \lim_{m \to 0}
\left\langle \frac{Z^m - 1}{m} \right\rangle_{\{ \sigma({\bf x}), \xi(n) \} } \, .
\end{equation}

First we average over the surface disorder $\sigma({\bf x})$.
Introducing $m$ replicas of polymer chains and making use of the properties of the
Gaussian distribution, Eq.~(\ref{surf_prob}), we obtain
\begin{widetext}
\begin{eqnarray}
\label{Z_repl_surface}
\langle Z^m \rangle_{\sigma({\bf x})}  &=&
\int \! \! \prod_{\alpha=1}^{m}\mathcal{D}\{{\bf r}_{\alpha}(n)\} \:
\exp\left[ -\frac{3}{2a^2}\sum_{\alpha=1}^{m}
\int_0^N \!\! dn\left(\frac{\partial {\bf r}_{\alpha}}{\partial n}\right)^2
\right]
\\ &&
\times \exp \left[ \:
\beta \sigma_0 \int_0^N \! dn \: \xi(n) \sum_{\alpha=1}^{m} V(z_{\alpha}(n))
+ \frac{1}{2}
\int \! \! \! \! \! \int_0^N \! \! dn_1 \: dn_2 \: \xi(n_1) \: \xi(n_2)
\: \widehat{g}(n_1,n_2) \right] \, ,
\nonumber
\end{eqnarray}
\end{widetext}
where $\alpha$ is the replica index and $\int \mathcal{D}\{{\bf
r}(n)\}$ denotes integration over all possible trajectories ${\bf
r}(n)$ of the chain. In Eq. (\ref{Z_repl_surface}) we have
introduced the notation
\begin{eqnarray}
\lefteqn{
\widehat{g}(n_1, n_2)=
} \\
&&
\beta^2 \sum_{\alpha,\gamma=1}^{m} 
V(z_{\alpha}(n_1)) \: V(z_{\gamma}(n_2)) \: g({\bf x}_{\alpha}(n_1)-{\bf x}_{\gamma} (n_2))
\nonumber
\end{eqnarray}
for clarity and future reference.

Next we average over the sequence distribution $\xi(n)$. This
affects only the second exponential in (\ref{Z_repl_surface}).
Again, exploiting the properties of Gaussian integrals, we obtain,
after some algebra (see appendix \ref{app:sequence_average})
\begin{widetext}
\begin{eqnarray}
\label{Z_repl_all}
\langle Z^m \rangle_{\{ \sigma({\bf x}), \xi(n) \} } & =&
\int \prod_{\alpha=1}^{m}\mathcal{D}\{{\bf r}_{\alpha}(n)\} \:
\exp\left[ -\frac{3}{2a^2}\sum_{\alpha=1}^{m} \int_0^N
dn\left(\frac{\partial {\bf r}_{\alpha}}{\partial n}\right)^2 \right] \:
\frac{1}{\sqrt{\det K}}
\\
&\times \exp & \bigg[
\sigma_0 \: \xi_0
\int \!\!\!\!\! \int_0^N \! dn_1 \: dn_2 \: K^{-1}(n_1,n_2) \:
\beta \sum_{\alpha=1}^{m} V(z_{\alpha}(n_2)) \bigg]
\nonumber\\
&\times \exp & \bigg[ \frac{\xi_0^2}{2}
\int \!\!\!\!\! \int \!\!\!\!\! \int_0^N \!\! dn_1 \: dn_2 \: dn_3 \: K^{-1}(n_1,n_2) \:
\widehat{g}(n_2,n_3) \bigg]
\nonumber\\
&\times \exp & \bigg[ \frac{\sigma_0^2}{2}
\int \!\!\!\!\! \int \!\!\!\!\! \int_0^N \!\! dn_1 \: dn_2 \: dn_3 \: K^{-1}(n_1,n_2) \:
c(n_3-n_1) \:
\beta^2 \sum_{\alpha, \gamma=1}^{m}  V(z_{\alpha}(n_2)) \: V(z_{\gamma}(n_3))
\bigg]  \, ,
\nonumber
\end{eqnarray}
\end{widetext}
where $K(n_1,n_2)$ is defined as
\begin{equation}
\label{kk}
K(n_1,n_2) = \delta(n_1-n_2)
- \int_0^N \!\!\! dn_3 \:
\widehat{g}(n_1,n_3) \: c(n_3 -n_2) \, .
\end{equation}

Eq.~(\ref{Z_repl_all}) covers the cases of homogeneous and heterogeneous polymers
adsorbing on homogeneous and heterogeneous surfaces. If at least one of the partners
is homogeneous, the matrix $K$ becomes unity, $K(n_1,n_2) = \delta(n_1-n_2)$, and
$\det K = 1$. If the polymer is homogeneous (a homopolymer), the argument of the
last exponential of (\ref{Z_repl_all}) vanishes and we recover Eq.~(\ref{Z_repl_surface})
in the special case $\xi(n) \equiv \xi_0$. If the surface is homogeneous, the argument of
the second last exponential in (\ref{Z_repl_all}) vanishes and we recover Eq.~(6)
of Ref.~\onlinecite{Polotsky:2003}.

So far the discussion has been fairly general. In the following, we shall mostly
restrict ourselves to two specific situations: Homopolymer adsorption on
heterogeneous surfaces ($c(n) \equiv 0, K(n_1,n_2) = \delta(n_1-n_2)$),
and the adsorption of overall neutral heteropolymers on overall neutral heterogeneous
surfaces ($\sigma_0 = \xi_0 = 0$). In the second case, the last three exponentials
in Eq.~(\ref{Z_repl_all}) can be dropped, and the effect of the disorder
comes in solely through the determinant $\det K \neq 1$.

\subsection{Definition of the Reference System}

To proceed with the calculations, we introduce a reference system \cite{Chen:2000}
with conformational and thermodynamic properties reasonably close to those of the
original system. A natural choice for the reference system is a single homopolymer
near an adsorbing homogeneous surface, interacting via an attractive potential
of the same form $V(z)$ as in (\ref{potential}), with the interaction strength
$w_0=\sigma_0 \: \xi_0 + v_0$. The parameter $v_0$ is a variational parameter
which can be adjusted such that the reference system is as close as possible to
the original one. The Hamiltonian of the reference system has the following form:
\begin{equation}
\label{Hamiltonian_ref} \frac{H_0}{kT}=\frac{3}{2a^2}\int_0^N
dn\left(\frac{\partial {\bf r}}{\partial n}\right)^2-\beta w_0
\int_0^N \, dn \, V(z(n)) \, .
\end{equation}

The next steps are very similar to those already described in our earlier
publication~\cite{Polotsky:2003} for the case of heteropolymer chain adsorption
on a homogeneous planar surface, and shall be only sketched very briefly here.
We begin with recasting the quantity $\langle Z^m \rangle$ from Eq.~(\ref{Z_repl_all})
in the form
\begin{equation}
\langle Z^m \rangle = Z_0^m \big[ \overline{\exp(A_m)}\big]_0^{(m)},
\end{equation}
where $Z_0$ is the partition function of the reference system, and
$[\overline{\cdots}]_0^{(m)}$ denotes the chain average with respect to $m$ independent
reference chains. The general expression for $A_m$ is complicated, but can be obtained from
Eq.~(\ref{Z_repl_all}) in a straightforward way.
In the case of homopolymers adsorbed on a heterogeneous surface,
the quantity $A_m$ is given by
\begin{equation}
\label{a_homo}
A_m = \frac{\xi_0^2}{2} \int \!\!\!\!\! \int_0^N dn_1 \: dn_2 \: \widehat{g}(n_1,n_2)
- \beta v_0 \int_0^N \! dn \sum_{\alpha=1}^m V(z_{\alpha}(n)) \,,
\end{equation}
and in the case of a neutral heteropolymer on a neutral surface, we get
\begin{equation}
\label{a_hetero}
A_m = - \frac{1}{2} \ln (\det K)
- \beta v_0 \int_0^N \! dn \sum_{\alpha=1}^m V(z_{\alpha}(n)) \,.
\end{equation}
The free energy difference between the original system and the reference system
is given by:
\begin{eqnarray}
\label{df}
\beta \Delta F &=& \lim_{m\to 0} (\langle Z^m \rangle - Z_0^m)/m
\\
 &=& \lim_{m \to 0}([\overline{\exp(A_m)}]_0^{(m)} - 1)/m \,.
\nonumber
\end{eqnarray}
Unfortunately, the expression $([\overline{\exp(A_m)}]_0^{(m)}-1)$
cannot be calculated exactly. Therefore, we expand it up to
leading order in powers of $v_0$, $\widehat{g}$, and $c$ (\ie
$\Delta_s$ and $\Delta_p$). In the homopolymer case,
Eq.~(\ref{a_homo}), this is equivalent to approximating $\exp(A)-1
\approx A$. In the heteropolymer case, Eq.~(\ref{a_hetero})) we
also have to expand $\ln (\det K)$:
\begin{eqnarray}
\label{exp_ln_det} \ln (\det K) &=& \Tr (\ln K) \approx \Tr (K -
{\bf 1}) 
\\ &=&
- \int \!\!\!\!\! \int_0^N \! dn_1 \: dn_2 \:
\widehat{g}(n_1,n_2) \: c(n_2-n_1)
\nonumber
\end{eqnarray}

The resulting expression for homopolymers on a heterogeneous surface reads
\begin{widetext}
\begin{eqnarray}
\label{df_homo}
\frac{\beta \Delta F}{N} &\approx& - \: \beta \: v_0 [\overline{ V(z(n)) }]_0
\\ &&
+ \: \beta^2 \: \xi_0^2 \int_0^N \!\! dk \int \! d{\bf x}' \: d{\bf x}'' \:
g({\bf x}'-{\bf x}'') \:
[\overline{ V(z(n)) \delta({\bf x}(n)-{\bf x}') V(z(n+k)) \delta({\bf x}(n+k)-{\bf x}'') }]_0
\nonumber \\ &&
- \:  \beta^2 \: \xi_0^2 \int \! d{\bf x}' \: d{\bf x}'' \:
g({\bf x}'-{\bf x}'') \:
[\overline{ V(z(n)) \delta({\bf x}(n)-{\bf x}') }]_0 \;
[\overline{ V(z(n)) \delta({\bf x}(n)-{\bf x}'') }]_0 \; N \,.
\nonumber
\end{eqnarray}
\end{widetext}
The last term corresponds to contributions from different replicas, whereas the other
two belong to single replicas. We have used the fact that averages for different replicas
factorize, since different replicated chains are independent in the reference system.
Thus $[\overline{\cdots }]_0$ denotes the average with respect to a single reference chain,
\parbox{0.48\textwidth}{
\begin{equation}
\label{average_def}
[\overline{X}]_0=
\frac {\int \mathcal{D}\{{\bf r}(n)\} \exp[-\beta H_0({\bf r}(n))] X({\bf r}(n))}
{\int \mathcal{D}\{{\bf r}(n)\} \exp[-\beta H_0({\bf r}(n))]} \,.
\end{equation}
}
Likewise, we obtain for neutral heteropolymers on neutral heterogeneous surfaces
\begin{widetext}
\begin{eqnarray}
\label{df_hetero}
\frac{\beta \Delta F}{N} &\approx& - \: \beta \: v_0 [\overline{V(z(n))}]_0
\\ &&
+ \: \beta^2 \int_0^N \!\! dk \; c(k) \int \! d{\bf x}' \: d{\bf x}'' \:
g({\bf x}'-{\bf x}'') \:
[\overline{ V(z(n)) \delta({\bf x}(n)-{\bf x}') V(z(n+k)) \delta({\bf x}(n+k)-{\bf x}'')}]_0
\nonumber \\ &&
- \: \beta^2 \int \! d{\bf x}' \: d{\bf x}'' \:
g({\bf x}'-{\bf x}'') \:
[\overline{ V(z(n)) \delta({\bf x}(n)-{\bf x}') }]_0 \;
[\overline{ V(z(n)) \delta({\bf x}(n)-{\bf x}'') }]_0 \; \int_0^N \! \! dk \; c(k)
\nonumber
\end{eqnarray}
\end{widetext}
As in Eq.~(\ref{df_homo}), the first two terms contain
contributions from the same replica, and the last term accounts
for contributions from different replicas.

At this point, the parameter $v_0$ is still a free variational
parameter of the theory. In the final step, we will choose it such
that the free energies of the reference system and the original
system are the same, i.e. $\Delta F = 0$.

\subsection{Green's Function and Transition Point of the Reference System}
To calculate the averages in (\ref{df_homo}) and (\ref{df_hetero}) we need to
find the Green's function of the reference system. It satisfies
the following differential equation \cite{DoiEdwards,deGennes_book}:
\begin{equation}
\label{diff_eq}
-\frac{\partial G}{\partial n} =
-\frac{a^2}{6}\nabla^2 G -\beta w_0 \delta(z-z_0) G({\bf r},{\bf r}';n) \, ,
\end{equation}
with the appropriate initial and boundary conditions:
\begin{eqnarray}
\label{initial_cond}
G({\bf r},{\bf r}';0) & = & \delta({\bf r}-{\bf r}') \, , \\
\label{boundary_cond_infty}
G({\bf r},{\bf r}';N)|_{|{\bf r}| \to \infty} &=& 0 \, , \\
\label{boundary_cond_zero}
G({\bf r},{\bf r}';N)|_{z=0} &=& 0 \, .
\end{eqnarray}
First, one can separate the variables $x$, $y$ and $z$ writing:
\begin{equation}
\label{sep_xyz} G({\bf r},{\bf r}';n)
=G_{\bf x}({\bf x},{\bf x}';n)\: G_z(z,z';n) \, ,
\end{equation}
$G_{\bf x}({\bf x},{\bf x}';n)$ is the free chain Gaussian
distribution in two dimensions:
\begin{equation}
\label{G_xy} G_{\bf x}({\bf x},{\bf x}';n)=\frac{3}{2\pi n a}
\exp\left[ \frac{3({\bf x}-{\bf x}')^2}{2na^2} \right] \, .
\end{equation}
For the $z$-dependent part of the Green's function (\ref{sep_xyz})
in the long-chain limit $n \gg 1$, one obtains in the ground-state
dominance approximation~\cite{Aslangul:1995, Polotsky:2003}
\begin{eqnarray}
\label{gs_G2}
\lefteqn{
G_{z}(z,z';n) = \frac{k_0/2}{1+(2k_b-k_0)z_0}
 \; e^{-\varepsilon_0 n} \,.
}
\\ &&
\left[ e^{-k_b|z-z_0|} - e^{-k_b(z+z_0)} \right]
 \; \left[e^{-k_b|z'-z_0|} - e^{-k_b(z'+z_0)} \right]
\nonumber
\end{eqnarray}
where $k_b$, determining the ground-state energy
($\varepsilon_0=k_b a^2/6$), is the root of the transcendental
equation
\begin{equation}
\label{k0_def}
\frac{2k}{1-e^{-2k z_0}}= k_0
\qquad \mbox{with} \qquad
k_0=\frac{6\beta w_0}{a^2} \,.
\end{equation}
The complete Green's function is calculated and discussed in the Appendix \ref{app:G}.

The adsorption transition in the reference system takes place when
$k_b$ and $\varepsilon_0$ vanish. This corresponds to the
condition~\cite{Polotsky:2003} $z_0 k_0=1$ or, according to the
definition (\ref{k0_def}) for $k_0$, to $6\beta w_0/a^2=1/z_0$.

\subsection{Free energy difference between the reference system
and the original system}
\label{deltaF}
With the results from the previous subsection, we can now evaluate
the averages in (\ref{df_homo}) and (\ref{df_hetero}). Using the
ground-state dominance approximation (\ref{gs_G2}) for the Green's
function, we obtain the same result for $[\overline{V(z(n))}]_0$
as in Ref. \onlinecite{Polotsky:2003}:
\begin{equation}
\label{V_av}
[\overline{V(z(n))}]_0 = \frac{2k_b^2/k_0}{1+(2k_b-k_0)z_0} \, .
\end{equation}
Correspondingly, the averages $[\overline{V(z(n)) \delta({\bf x}(n)-{\bf x}')}]_0$
appearing in the last term of (\ref{df_homo}) and (\ref{df_hetero}) are given by
\begin{equation}
\label{V2_av_diff}
\big[\overline{V(z(n)) \delta({\bf x}(n)-{\bf x}')}\big]_0
= \frac{1}{S} \left[\frac{2k_b^2/k_0}{1+(2k_b-k_0)z_0}\right] \,.
\end{equation}
where $S$ is the surface area. The additional factor $1/S$ in
(\ref{V2_av_diff}), compared to (\ref{V_av}), results from the
translational freedom of the polymer in the surface plane.
Substituting this result into the last term of (\ref{df_homo}) and
(\ref{df_hetero}) and performing the integration over $\int d{\bf
x}' \: d{\bf x}''$ gives the factor $1/S$. Since $S \to
\infty$, this contribution is infinitesimally small and can be
omitted in the following calculations.

This seemingly ``technical'' point has an important physical
interpretation. After dropping the last term, Eqs.~(\ref{df_homo})
and (\ref{df_hetero}) do no longer contain terms with
contributions from different replicas. These terms, however, are
the only ones which distinguish between {\em quenched} and {\em
annealed} disorder. One can see that by directly carrying out an
analogous calculation for annealed disorder. In the annealed case, one
has to average directly over the partition function and calculate
\begin{equation}
 \beta \delta F_{\mbox{\tiny annealed}}
= \ln \langle Z \rangle -\ln Z_0
= \ln [\overline{\exp(A_1)}]_0.
\end{equation}
After expanding up to leading order in powers of $v_0$, $\widehat{g}$ and $c$,
one obtains the same expressions as (\ref{df_homo}) and (\ref{df_hetero}),
except that the last term is missing.

The assumption that quenched surface disorder is equivalent to
annealed surface disorder has been used in replica mean field
calculations~\cite{Srebnik:1996} and Monte Carlo
simulations~\cite{Bratko:1998}, based on the argument that the
chain ``samples all significant surface patterns with a
probability that asymptotically approaches that of an annealed
system in the adiabatic limit'' (cited from Ref.
\onlinecite{Bratko:1998}). Here we recover this equivalence in the
limit of infinite surface area $S$.

We turn to the calculation of $[\overline{\delta({\bf x}(n+k)-{\bf
x}') V(z(n+k)) \delta({\bf x}(n)-{\bf x}'')V(z(n))}]_0$. We assume
the validity of the ground-state dominance approximation for the
overall chain and for the side subchains. This gives
\begin{widetext}
\begin{eqnarray}
\label{V2_av_same}
\lefteqn{
 [\overline{\delta({\bf x}(n+k)-{\bf x}') V(z(n+k)) \delta({\bf x}(n)-{\bf x}'')V(z(n)) }]_0
} \qquad \qquad  &&
\\
  & = & \frac{\int d{\bf r}_1 \, d{\bf r}_2 \, d{\bf r}_3 \, d{\bf r}_4 \,
G({\bf r}_1, {\bf r}_2;n) \delta({\bf r}_2-{{\bf x}' \choose z_0}
) G({\bf r}_2, {\bf r}_3;k)\delta({\bf r}_3-{{\bf x}'' \choose
z_0}) G({\bf r}_3,{\bf r}_4;N-n-k) } {\int d{\bf r}_1 \, d{\bf
r}_2 \, d{\bf r}_3 \, d{\bf r}_4 \, G({\bf r}_1, {\bf r}_2;n)
G({\bf r}_2, {\bf r}_3;k) G({\bf r}_3,{\bf r}_4;N-n-k) }
\nonumber \\
&= & \frac{1}{S}G_{\bf x}({\bf x}', {\bf x}''; k)G_z(z_0,z_0; k)
e^{-|\varepsilon_0| k}\frac{2k_b^2/k_0}{1+(2k_b-k_0)z_0}
\nonumber
\end{eqnarray}
\end{widetext}

We consider first the homopolymer case.
Inserting (\ref{V_av}) -- (\ref{V2_av_same}) into (\ref{df_homo}), we obtain
\begin{eqnarray}
\label{df_homo2}
\lefteqn{
\frac{\beta \Delta F}{N} =
\frac{2k_b^2/k_0}{1+(2k_b-k_0)z_0} 
} \\&&
\left[ \frac{\beta^2 \xi_0^2}{S}
\int d{\bf x} \, d{\bf x}' \; g({\bf x}-{\bf x}') \;
\mathcal{G}({\bf x}, z_0, {\bf x}', z_0,  |\varepsilon_0|)|
- \beta v_0 \right] .
\nonumber
\end{eqnarray}
Here
\begin{eqnarray}
\mathcal{G}({\bf r}, {\bf r}', p)& =& \int_0^N dk \, e^{-pk}
G({\bf r}, {\bf r}'; k)
\\ 
&\simeq& \int_0^\infty dk \, e^{-pk}
G({\bf r}, {\bf r}'; k)
\nonumber
\end{eqnarray}
is the Laplace transform of the "complete" Green's function
$G({\bf r}, {\bf r}'; k)$. We are particularly interested in
its value at $z=z'=z_0$ ({\em i.~e.}, at the level where the potential
acts on the chain). The procedure of calculating $\mathcal{G}
({\bf x}, z_0,{\bf x}', z_0; |\varepsilon_0|)$ is briefly described in
the Appendix \ref{app:G}. The result can be written in the
following integral form
\begin{eqnarray}
\label{g1_z0_z0_rep}
\lefteqn{
\mathcal{G}({\bf x},  z_0,{\bf x}', z_0; |\varepsilon_0|)=
\frac{6}{a^2} \frac{1}{(2\pi)^2} \int d{\bf q} \; e^{-i{\bf q}({\bf x}-{\bf x}')}
}\\&&
\times
\frac{\sinh \left(z_0\sqrt{q^2 + k_b^2}\right)}{\sqrt{q^2 + k_b^2}
e^{z_0\sqrt{q^2 + k_b^2}}-k_0\sinh\left(z_0\sqrt{q^2 + k_b^2}\right)}
\nonumber
\end{eqnarray}
where ${\bf q}$ is the two-dimensional vector and
$q \equiv |{\bf q}|$. Inserting (\ref{g1_z0_z0_rep}) into (\ref{df_homo2})
and changing the order of integration over ${\bf q}$ and ${\bf x}$ yields
the following expression for the free energy difference between the
reference system and a homopolymer adsorbed on a heterogeneous surface:
\begin{widetext}
\begin{equation}
\label{df_homo_final}
 \frac{\beta \Delta F}{N} = \frac{2k_b^2/k_0}{1+(2k_b-k_0)z_0} \\
\left[ \frac{\beta^2\Delta_s^2 \xi_0^2}{a^2}\int_0^\infty \!\! dq \:
\frac{\Gamma_s \: q}{\left( q^2 + \Gamma_s^2\right)^{3/2}}
\frac{6 \sinh \left(z_0\sqrt{q^2 + k_b^2}\right)}
 {\left( \sqrt{q^2 + k_b^2} e^{z_0\sqrt{q^2 + k_b^2}}-k_0\sinh(z_0\sqrt{q^2 + k_b^2})\right)}
- \beta v_0 \right]
\end{equation}
\end{widetext}

The second case of interest is the neutral heteropolymer on
the neutral heterogeneous surface. The difference between the Eqs.~(\ref{df_homo})
and (\ref{df_hetero}) is due to the monomer-monomer correlation function
(\ref{seq_corrf}). Inserting expressions (\ref{V_av} - \ref{V2_av_same}) into
(\ref{df_hetero}), we obtain
\begin{widetext}
\begin{equation}
\label{df_hetero2}
\frac{\beta \Delta F}{N} = \frac{2k_b^2/k_0}{1+(2k_b-k_0)z_0}
\left[ \frac{\beta^2 \Delta_p^2}{S} \int d{\bf x} \, d{\bf x}' \;
g({\bf x} - {\bf x}') \; \mathcal{G}({\bf x}, z_0, {\bf x}', z_0,
\Gamma_p+|\varepsilon_0|) - \beta v_0 \right] .
\end{equation}
This finally leads to
\begin{equation}
\label{df_hetero_final}
 \frac{\beta \Delta F}{N} = \frac{2k_b^2/k_0}{1+(2k_b-k_0)z_0} 
\left[ \frac{\beta^2\Delta_s^2 \Delta_p^2}{a^2}\int_0^\infty \!\! dq \:
\frac{\Gamma_s \: q}{\left( q^2 + \Gamma_s^2\right)^{3/2}}
\frac{6 \sinh \left(z_0 \: \omega(q) \right)}
 {\left( \omega(q) \: e^{z_0\omega(q)}-k_0\sinh(z_0 \: \omega(q))\right)}
- \beta v_0 \right],
\end{equation}
\end{widetext}
with $\omega(q) \equiv \sqrt{q^2 + 6\Gamma_p/a^2 + k_b^2}$. Note
that (\ref{df_homo_final}) can formally be obtained from
(\ref{df_hetero_final}) by setting $\Delta_p=1$ and  $\Gamma_p \to
0$. This is, however, not a general rule. The corresponding
expression for a heteropolymer adsorbed on a homogeneous surface,
Eq.~(41) in Ref.~\onlinecite{Polotsky:2003} is not fully contained
in Eq.~(\ref{df_hetero_final})~\cite{footnote}

\subsection{Transition point}
\label{Transition}

Having obtained an approximate expression for the difference $\Delta F$
between the free energies of the original and the reference systems,
we adjust the auxiliary parameter $v_0$ such that the free energy of the
reference system is the best approximation for that of the original one.
Thus we require $ \Delta F(v_0^*)=0 $. For the $\Delta F$ given by (\ref{df_homo_final})
or (\ref{df_hetero_final}), this equation has one trivial solution $k_b=0$ or $w_0=0$,
corresponding to the uninteresting case of a chain which is desorbed in both the original
and the reference system.
For the other, nontrivial, solution we again consider both cases separately.

In the homopolymer case, the nontrivial solution for $v_0$ is given by setting the term
in square brackets in Eq.~(\ref{df_homo_final}) to zero.
The transition point for the reference system corresponds to the
condition ~\cite{Polotsky:2003} $z_0 k_0=1$ and $k_b \to 0$. Substituting these conditions
and introducing, for the sake of convenience, the dimensionless variables
\begin{displaymath}
\widetilde{v_0} = \frac{\beta v_0}{a}, \quad
\widetilde{\sigma_0} = \frac{\beta \sigma_0}{a}, \quad
\widetilde{\Delta_s} = \frac{\beta \Delta_s}{a},\quad
\end{displaymath}
\begin{equation}
\widetilde{\Gamma_s} = a\Gamma_s, \mbox{ and} \quad
\widetilde{z_0} = \frac{z_0}{a}
\label{dimless}
\end{equation}
we obtain an equation for the transition point $\widetilde{\sigma_0} \xi_0)^{tr}$
in the random system:
\begin{widetext}
\begin{equation}
\label{trp_homo}
(\widetilde{\sigma_0} \xi_0)^{tr} - (\widetilde{\sigma_0} \xi_0)_{\mbox{\tiny homo}}^{tr}=
- \beta \widetilde{v_0} 
=
- 6 \: \widetilde{\Delta_s}^2 \xi_0^2 \int_0^\infty \!\! dq \:
\frac{\widetilde{\Gamma_s}}{\left( q^2 + \widetilde{\Gamma_s}^2\right)^{3/2}} \;
\frac{\sinh\left(\widetilde{z_0} q\right)}
{\left(e^{\widetilde{z_0} q} - \sinh(\widetilde{z_0} q)/(\widetilde{z_0} q)\right)} \;,
\end{equation}
\end{widetext}
where $(\widetilde{\sigma_0} \xi_0)^{tr}_{\mbox{\tiny homo}} \equiv 1/(6\widetilde{z_0})$
is the transition point for the homopolymer near an attractive
surface. Eq.~(\ref{trp_homo}) determines the relation between
$\sigma$, $\Gamma_s$, and $\Delta_s$ that correspond to the
adsorption-desorption transition in the original heterogeneous system.

We turn to the case of a neutral heteropolymer on a neutral heterogeneous surface.
Here, the nontrivial solution for $v_0$, is obtained by setting the term
in square brackets in Eq.~(\ref{df_hetero_final}) to zero.
We insert again the conditions for the transition point in the reference system,
$k_b=0$ and $k_0 \: z_0 = 1$, and switch to the dimensionless variables according
to Eq.~ (\ref{dimless}). The parameters $\Gamma_p$ and $\Delta_p$ associated
with the RHP statistics are already dimensionless. Since $\widetilde{\sigma_0}=\xi_0=0$,
the equation for the transition point then reads:
\begin{widetext}
\begin{equation}
\label{trp_hetero}
\beta \widetilde{v_0} =
(\widetilde{\sigma_0} \xi_0)_{\mbox{\tiny homo}}^{tr}=
\int_0^\infty \!\!\! dq \: \frac{ 6 \: \sinh\left(\widetilde{z_0} \sqrt{q^2+6\Gamma_p}\right)}
{\left(\sqrt{q^2+6\Gamma_p} \; e^{\widetilde{z_0} \sqrt{q^2+6\Gamma_p}} -
 \sinh\left(\widetilde{z_0} \sqrt{q^2+6\Gamma_p}\right)/\widetilde{z_0}\right) } \;
\frac{\widetilde{\Gamma_s} \: q \: \widetilde{\Delta_s}^2 \Delta_p^2}
     {\left( q^2 + \widetilde{\Gamma_s}^2\right)^{3/2}}.
\end{equation}
\end{widetext}

\section{Summary and Discussion}
\label{Conclusions}

We have used a reference system approach to calculate the
adsorption/desorption transition of polymers at surfaces in systems with spatially
correlated randomness. The resulting expressions become very elegant in the limit
$\widetilde{z_0} \ll 1$, \ie if the range of the surface
potential is short compared to the statistical segment length of
the polymer. We shall now compile these results, contrast them with the
corresponding result in our earlier paper~\onlinecite{Polotsky:2003},
and discuss them.

We briefly recall the meaning of the different parameters. $\xi_0$
denotes the mean polymer charge per segment and $\widetilde{\sigma_0}$ the
mean surface charge per site; thus the mean affinity between polymer segments
and surface sites is given by the parameter $Q=(\widetilde{\sigma_0} \xi_0)$.
Randomness shifts the critical affinity parameter at the adsorption transition
$Q^{(tr)}$ to lower values, compared to the corresponding value
$Q^{(tr)}_{\mbox{\tiny Homo}}$ of a homogeneous system.
The shift depends on the variances $\Delta_p$ and $\widetilde{\Delta_s}$
of the polymer and surface disorder, and on the cluster parameters
or inverse correlation lengths $\Gamma_p$ and $\widetilde{\Gamma_s}$.

The case of random heteropolymers on homogeneous surfaces was already
considered in our earlier paper~\cite{Polotsky:2003}. The transition
in the random system is shifted with respect to the homogeneous system by
\begin{equation}
\label{trp_rhp_final}
Q^{(tr)}-Q^{(tr)}_{\mbox{\tiny Homo}} =
- \Delta_p^2 \widetilde{\sigma_0}^2 \sqrt{\frac{6}{\Gamma_p}}.
\end{equation}
The case of homopolymers on random surfaces is covered by Eq.~(\ref{trp_homo}).
In the limit $z_0 \ll 1$, the integral in (\ref{trp_homo}) can be calculated
explicitly and the solution takes the form
\begin{equation}
\label{trp_rhs_final}
Q^{(tr)}-Q^{(tr)}_{\mbox{\tiny Homo}} =
- \widetilde{\Delta_s}^2 \xi_0^2 \frac{6}{\widetilde{\Gamma_s}}.
\end{equation}
Heterogeneous polymers on heterogeneous surfaces were discussed in the special case
of zero mean affinity. The result, given by Eq.~(\ref{trp_hetero}),
simplifies considerably in the limit $z_0 \ll 1$.
The heteropolymer is adsorbed, if
\begin{equation}
\label{trp_hetero_final}
6 \frac{\widetilde{\Delta_s}^2 \Delta_p^2}{\widetilde{\Gamma_s}+\sqrt{6\Gamma_p}}
\ge Q_{\mbox{\tiny Homo}}^{tr} = \mbox{const.}
\end{equation}
This defines a phase boundary in the parameter space of variances $\Delta_p$ and
$\widetilde{\Delta_s}$ and cluster parameters $\Gamma_p$ and $\widetilde{\Gamma_s}$.

In order to complete this compilation of results, we also give the
general expression for the shift of the adsorption transition in
systems of heterogenous on heterogeneous surfaces (not necessarily
neutral). It contains the sum of the three contributions
(\ref{trp_rhp_final}), (\ref{trp_rhs_final}),
(\ref{trp_hetero_final}), but has one extra term from the
expansion of $K^{-1} = \sum_{k=0}^{\infty} (\int_0^N \! dn_1 \:
\widehat{g}(n_1,n_2) \: c(n_2-n_1))^k$ in (\ref{Z_repl_all}). The
calculation is absolutely analogous to the one presented in
Sec.~(\ref{reference}) and yields
\begin{eqnarray}
\label{trp_all_final}
Q^{(tr)}-Q^{(tr)}_{\mbox{\tiny Homo}} &=&
- \Delta_p^2 \widetilde{\sigma_0}^2 \sqrt{\frac{6}{\Gamma_p}}
- \widetilde{\Delta_s}^2 \xi_0^2 \frac{6}{\widetilde{\Gamma_s}}
\\&&
- \widetilde{\Delta_s}^2 \Delta_p^2 \: C  + {\cal O}(\widetilde{\Delta_s}^4,\Delta_p^4)
\nonumber
\end{eqnarray}
where the coefficient $C$ of the higher order term  $\widetilde{\Delta_s}^2 \Delta_p^2$ is given by
\begin{eqnarray}
\label{C_coeff}
C &=&
\frac{6}{\widetilde{\Gamma_s}+\sqrt{6\Gamma_p}} - 72 \:
\widetilde{\sigma_0} \xi_0 
\left\{ \frac{1}{\sqrt{6 \Gamma_p}}
\Big(
  \frac{1}{\widetilde{\Gamma_s}} +
  \frac{1}{\widetilde{\Gamma_s} + \sqrt{6 \Gamma_p}} \Big)
\right.
\nonumber
\\&& \qquad 
\left.
+ \frac{1}{\pi \widetilde{\Gamma_s}^2} G^{22}_{22} \Big(
\frac{\widetilde{\Gamma_s}^2}{6 \Gamma_p} \Big|
\begin{array}{cc}1&1\\\frac{1}{2}&\frac{3}{2} \end{array} \Big)
\right\},
\end{eqnarray}
with the Meijer G function $G$. The calculation of $C$ is shortly
sketched in the Appendix \ref{app:all}. Note that Eq.~(\ref{trp_all_final})is 
an implicit equation for $Q^{(tr)}$, because the affinity $Q=
(\widetilde{\sigma_0} \xi_0)$ depends both on the mean charge per
segment $\xi_0$ of the polymer and the mean charge
$\widetilde{\sigma_0}$ per site on the surface.

\begin{figure}[t]
\begin{center}
\includegraphics[width=10cm,angle =0]{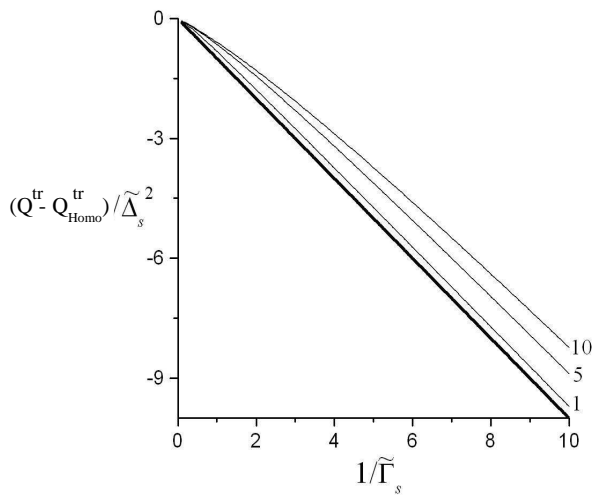}
\end{center}
\caption{ The shift
$(Q^{tr} - Q_{\mbox{\tiny Homo}}^{tr})/\widetilde{\Delta_s}^2$
of the adsorption-desorption transition point for a random surface with
exponentially decaying sequence correlations compared to that of a
homogeneous attracting surface, vs. correlation length
$1/\widetilde{\Gamma_s}$, according to Eq.~(\ref{trp_homo}).
The different curves show the results for different values of
$\widetilde{z_0}$ as indicated. The bold line corresponds to
$\widetilde{z_0}=0$, Eq.~(\ref{trp_rhs_final}).
\label{fig_exp} }
\end{figure}

The validity of the approximation $\widetilde{z_0} \ll 1$ is demonstrated in Fig.~\ref{fig_exp},
which shows some adsorption-desorption transition curves for homopolymers adsorbed
on heterogeneous surfaces at different values of $\widetilde{z_0}$, as calculated
from Eq.~(\ref{trp_homo}).
As a function of the correlation length $1/\Gamma_s$, the curves become
straight lines with increasing $1/\Gamma$ with the same slope as the
asymptotic curve, Eq.~(\ref{trp_rhs_final}). The slope of the
curves deviates from the asymptotic value only at small correlation
lengths. The simplified expression (\ref{trp_rhs_final})
approximates the full solution (\ref{trp_homo}) reasonably for
$\widetilde{z_0} < 1$. Thus the approximation is justified if the range of the
potential is shorter than the statistical segment length. At first sight,
this condition seems somewhat arbitrary, given that the value of the statistical
segment length $a$ depends on the particular choice of the ``segment'' unit.
One can always increase $a$ by choosing a more coarse grained polymer description,
\ie combining several segments to one new effective segment.
The real length scale set by the non-adsorbed polymer is its radius of gyration.
Applying the approximation $\widetilde{z_0} \ll 1$ in the calculation of the
adsorption transition is thus justified whenever the range of the potential
is much shorter than the radius of gyration of the polymer.

Our results can be summarized as follows: The presence of
correlations always enhances adsorption.  Both increasing the
correlation lengths, \ie decreasing in $\Gamma_p$ or $\Gamma_s$,
and increasing the strength of the correlations (given by
$\Delta_p$, $\Delta_s$) results in a shift of the adsorption
transition towards lower affinities.

The results obtained for homopolymers on random surfaces (\ref{trp_rhs_final})
and those for heteropolymers on homogeneous surfaces (\ref{trp_rhp_final}) are
very similar. The only difference are the exponents of $\Gamma_s$ and $\Gamma_p$,
which reflect the different "dimensions" of a planar surface and a linear
polymer chain.

The final result (\ref{trp_hetero_final}) for the adsorption of
neutral heteropolymers on neutral heterogeneous surfaces gives
some particularly interesting information. First, it is worth
noting that in Eq.~(\ref{trp_hetero_final}), both $\Gamma_s$ and
$\Gamma_p$ enter in the denominator with their "characteristic"
exponents (1 and 1/2, respectively). The other notable feature is
that $\Delta_s$ and $\Delta_p$ enter in (\ref{trp_hetero_final})
as the product of their squares $\widetilde{\Delta_s}^2
\Delta_p^2$ -- a similar dependence on the variances has been
observed by Srebnik \etal~\cite{Srebnik:1996} for the case of
short-range correlated RHPs at a random surface. This means that
if the surface (heteropolymer) contains attractive, repelling, and
neutral sites and is kept neutral on average, it is preferable to
have strongly repelling and strongly adsorbing sites rather than
weakly repelling and weakly adsorbing - "almost neutral" - sites.

We note another interesting property of the considered model. In
Sec.~\ref{deltaF}, we have already discussed the equivalence of
quenched and annealed {\em surface} disorder. The argument was
based on the fact that all terms related to different replicas in
Eqs.~(\ref{df_homo}) and (\ref{df_hetero}) contain an extra factor
$1/S$, which vanishes in the limit of infinite surface area $S$.
Technically, the factor $1/S$ reflects the loss of translational
freedom of a replica, if it is restricted to stay close to another
replica by a coupling through the correlation function $g({\bf
x}-{\bf x}')$. The argument is thus only valid for surface
disorder. In general, quenched and annealed disorder on the {\em
polymer} are not equivalent.

At the transition point, however, terms related to different replicas always
vanish, because every independent replica also contributes a factor
$2 k_b^2/k_0/(1 + (2 k_b-k_0) z_0) \sim 2 k_b$. In other words, our model calculation
predicts that the adsorption threshold is the same for quenched and annealed
surface disorder. We believe that this is a general feature of the adsorption transition
in disordered systems. Joanny finds the same equivalence when studying the adsorption
of polyampholytes on surfaces~\cite{Joanny:1994} within the Hartree approximation.
He claims that it has been conjectured for similar problems by
Nieuwenheuizen~\cite{Nieuwenheuizen}. Indeed, polymers and surfaces have only
very few contacts at the adsorption transition. It is conceivable that the
contacts can be arranged in an optimal way regardless of the particular type of
disorder.

If this is true, it has obvious implications for the phenomenon of molecular
recognition: It means that studies of the adsorption transition in systems with
quenched disorder, averaged with general, unspecific site distribution functions
such as Eqs.~(\ref{seq_prob}) or (\ref{surf_prob}), can only provide very limited
information about specific recognition processes. The adsorption threshold of an
annealed polymer, which has had the opportunity to adapt its sequence to the surface,
is absolutely equivalent to that of a completely unadapted (quenched) polymer.

\begin{figure}[t]
\begin{center}
\includegraphics[width=6cm,angle =0]{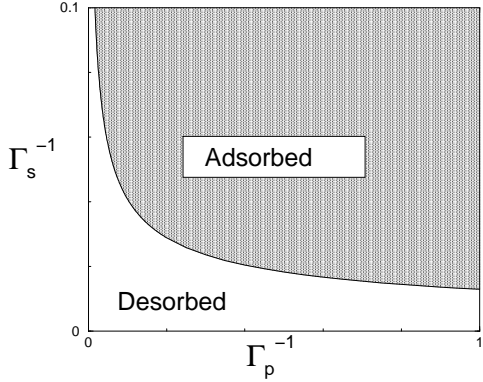}
\end{center}
\caption{
Adsorption transition of neutral heteropolymers on neutral surfaces
as a function of $\Gamma_p$ and $\widetilde{\Gamma_s}$. The parameters
are $\widetilde{\Delta_s} = \Delta_p =1$ and $Q_{\mbox{\tiny Homo}}^{(tr)}=1$.
\label{fig:adsorption} }
\end{figure}

We can also use our results to discuss the issue of the interplay between cluster
sizes on the surface and the homopolymer. Fig.~\ref{fig:adsorption} shows an
example of an adsorption phase diagram, obtained from Eq.~(\ref{trp_hetero_final}),
for neutral heteropolymers on neutral surfaces as a function of the cluster parameters
$\Gamma_p$ and $\widetilde{\Gamma_s}$. The phase boundary does not reflect particular
affinities between certain cluster parameters $\Gamma_p$ on the polymer and
$\widetilde{\Gamma_s}$ on the surface. Regardless of the value of $\Gamma_p$,
the adsorption is always highest at low $\widetilde{\Gamma_s}$ and vice versa.
We do not observe any complementarity effects.

To summarize, we have been able to derive simple and elegant equations for
the adsorption threshold in systems of polymers on surfaces with correlated
quenched and annealed disorder. These may be useful for future studies.
We have discussed our results in the context of molecular recognition and found
that the complex process of pattern recognition between molecules and surfaces
close to the adsorption threshold cannot be understood in terms of a simple model with
quenched or annealed disorder. We conclude that recognition either takes place
at higher affinities, beyond the adsorption threshold, as suggested by
Chakraborty \etal~\cite{Chakraborty_rev}, or has to be studied by more
sophisticated models and methods. For example, the situation might change
if the probability distribution itself reflects the pattern on the surface.
Such considerations shall be the subject of future study.

\section*{Acknowlegdement}
\label{Acknowlegdement}

The financial support of the Deutsche Forschungsgemeinschaft (SFB
613) is gratefully acknow\-ledged. A.P. thanks RFBR (grant
02-03-33127).

\appendix

\section{\label{app:sequence_average} Performing the Sequence Average}
We want to calculate surface disorder average in Eq
(\ref{Z_repl_surface}). Let us define, for convenience,
\begin{displaymath}
\bar{\sigma}(n) = \beta \sigma_0 \sum_{\alpha=1}^m V(z_{\alpha}(n))\,, \qquad
\bar{\xi}(n) \equiv \xi_0
\end{displaymath}
and adopt the matrix notation
\begin{displaymath}
\xi G \xi := \int \!\!\!\!\! \int_0^N dn_1 \: dn_2 \: \xi(n_1) \:
\end{displaymath}
\begin{displaymath}
G(n_1,n_2) \: \xi(n_2) \qquad \xi v := \int_0^N dn \: \xi(n) v(n)
\,.
\end{displaymath}
The matrices $c$, $c^{-1}$, and $\widehat{g}$ are symmetric.
In order to calculate the sequence average, we need to combine
Eqs.~(\ref{Z_repl_surface}) and (\ref{seq_prob}) and calculate
\begin{eqnarray}
I &= &\frac{ \int \mathcal{D}\{\xi(n)\} \: \exp(\xi \bar{\sigma} + \frac{1}{2} \xi \widehat{g} \xi )
     \exp( - \frac{1}{2} (\xi - \bar{\xi}) c^{-1} (\xi - \bar{\xi}) ) }
    { \int \mathcal{D}\{\xi(n)\} \: \exp(- \frac{1}{2} \xi c^{-1} \xi ) }
\nonumber\\
&=& \exp \left[- \frac{1}{2} \bar{\xi} c^{-1} \bar{\xi}
     + \frac{1}{2} (\bar{\sigma} + \bar{\xi} c^{-1})
           [c^{-1} - \widehat{g}]^{-1} (\bar{\sigma} + c^{-1} \bar{\xi}) \right]
\nonumber\\ && \times
\frac
{ \int \mathcal{D}\{\xi(n)\} \: e^{ - \frac{1}{2} \xi [c^{-1}-\widehat{g}] \xi} }
{ \int \mathcal{D}\{\xi(n)\} \: e^{- \frac{1}{2} \xi c^{-1} \xi} }
\label{a2}
\end{eqnarray}
The denominator corresponds to the normalization factor in
Eq.~(\ref{seq_prob}), and the expression $\int
\mathcal{D}\{\xi(n)\}$ denotes functional integration in sequence
space. To simplify the first exponential on the right hand side of
Eq.~(\ref{a2}), we define $K = {\bf 1} - \widehat{g} c$ (Eq.
(\ref{kk})) and replace $[c^{-1} - \widehat{g}]^{-1} = c K^{-1}$
thus obtain the expression
$
\exp\left[ \bar{\xi} K^{-1} \widehat{g} \bar{\xi}/2
+ \bar{\sigma} c K^{-1} \bar{\sigma}/2
+ \bar{\xi} K^{-1} \bar{\sigma} \right]
$.
This corresponds to the last three exponentials in Eq.~(\ref{Z_repl_all}).

To calculate the integrals over sequences $\{\xi(n)\}$ in Eq.~(\ref{a2}),
we set $n$ back to be a discrete variable,
\mbox{$\int \mathcal{D}\{\xi(n)\} = \int_0^{\infty} d\xi_1 \cdots \int_0^\infty d\xi_N$}.
This gives
\begin{eqnarray}
\lefteqn{
\frac
{ \int \mathcal{D}\{\xi(n)\} \: e^{ - \frac{1}{2} \xi [c^{-1}-\widehat{g}] \xi} }
{ \int \mathcal{D}\{\xi(n)\} \: e^{- \frac{1}{2} \xi c^{-1} \xi} }
=
\sqrt{\frac{\det(c^{-1})}{\det(c^{-1}-\widehat{g})}}
}\qquad  \\  \nonumber
 &=&
\frac{1}{\sqrt{\det[(c^{-1}-\widehat{g})c]}}=\frac{1}{\sqrt{\det(K)}} \,.
\end{eqnarray}
The final result, which enters Eq.~(\ref{Z_repl_all}), thus reads
\begin{equation}
I = \frac{1}{\sqrt{\det K}} \:
\exp \left[ \frac{1}{2} \bar{\xi} K^{-1} \widehat{g} \bar{\xi}
+ \frac{1}{2} \bar{\sigma} c K^{-1} \bar{\sigma}
+ \bar{\xi} K^{-1} \bar{\sigma} \right] .
\end{equation}

In matrix notation, it is easy to see that
\begin{eqnarray}
\label{trk} \ln (\det(K)) &=& \ln (\det(T K T^{-1})) 
\\ \nonumber
&=& \ln(\prod_n
\lambda_n) = \sum_n \ln (\lambda_n) = \Tr(\ln K) \,,
\end{eqnarray}
where $T$ diagonalizes $K$, and $\lambda_n$ are the eigenvalues of $K$.

\section{\label{app:G} Complete Green's Function}

Our starting point is the equation (\ref{diff_eq})
with the initial condition (\ref{initial_cond}) and the boundary
conditions ((\ref{boundary_cond_infty}) and (\ref{boundary_cond_zero})):
Laplace transforming with respect to $n$ and taking into account the initial condition
(\ref{initial_cond}), we obtain
\begin{eqnarray}
\label{Lapl1}
\lefteqn{
-p \: \mathcal{G}({\bf x}, z,{\bf x}', z';p)
+ \delta({\bf x}-{\bf x}')\delta(z-z')
} \\&& \nonumber
= -\frac{a^2}{6}\nabla^2 \mathcal{G}
- \beta \: w_0 \: \delta(z-z_0) \mathcal{G}({\bf x}, z,{\bf x}', z';p) \,,
\end{eqnarray}
where $\mathcal{G}({\bf x}, z,{\bf x}', z';p)$ is the Laplace transform of
$G({\bf x}, z,{\bf x}', z';n)$ with respect to $n$ with the boundary conditions
\begin{equation}
\label{Lapl1_bc}
\begin{split}
\mathcal{G}({\bf x}, z,{\bf x}', z';p)|_{|{\bf x}|, z \to \infty}&=0, \quad
\\ 
\mathcal{G}({\bf x}, z,{\bf x}', z';p)|_{z \to 0}&=0
\end{split}
\end{equation}
Firstly, we apply to this equation a Fourier transform with
respect to ${\bf x}$ (${\bf x}\to {\bf q}$ and $\mathcal{G}\to
\mathscr{G}$)
\begin{eqnarray}
\label{Four2}
\lefteqn{
-p \: \mathscr{G}({\bf q}, z,{\bf x}', z';p)  + e^{-{\bf q}{\bf x}}\delta(z-z')
} \\ \nonumber
&=& -\frac{a^2}{6} \left(-|{\bf q}|^2 \mathscr{G} +
\frac{\partial^2 \mathscr{G}}{\partial z^2} \right)
\\&& \nonumber
-\beta w_0 \delta(z-z_0)
\mathscr{G}({\bf q}, z,{\bf x}', z';p)
\end{eqnarray}
and secondly a Laplace transform with respect to $z$ ($z\to s$ and
$\mathscr{G} \to g$).
\parbox{0.45\textwidth}{
\begin{eqnarray}
\label{Lapl3}
\lefteqn{
-p \: g({\bf q}, s,{\bf x}', z';p) + e^{-{\bf q}{\bf x}'} e^{-z's}=
}\qquad  \\ &&
 - \: \frac{a^2}{6} \left[ \left( -|{\bf q}|^2 + s^2 \right) g({\bf q}, s,{\bf x}', z';p)
\right. 
\\ && \nonumber \qquad \qquad
\left.
- s \: \mathscr{G}({\bf q}, 0,{\bf x}', z';p)- \mathscr{G}'({\bf q}, 0,{\bf x}', z';p)
\right]
\nonumber \\ &&
 - \: \beta \: w_0 \: e^{-z_0 s} g({\bf q}, z_0,{\bf x}', z';p)
\nonumber
\end{eqnarray}
}
Here $\mathscr{G}'$ denotes $\partial \mathscr{G}/ \partial z$.
This equation can be solved for $g({\bf q}, s,{\bf x}', z';p)$, giving
\begin{widetext}
\begin{equation}
\label{g2}
g({\bf q}, s,{\bf x}', z';p) =
\frac{\mathscr{G}'({\bf q}, 0,{\bf x}', z';p)
-(6\beta w_0/a^2) \: \mathscr{G}({\bf q}, z_0,{\bf x}', z';p)
\: e^{-z_0 s}-(6/a^2)e^{-z' s}
e^{-{\bf q}{\bf x}'}}{s^2-|{\bf q}|^2-6p/a^2} \,.
\end{equation}

Taking the inverse Laplace transform of (\ref{g2}) with respect to $s$ and the inverse
Fourier transform with respect to ${\bf q}$, we obtain for
$\mathcal{G}({\bf x}, z_0, {\bf x}', z_0; p)$:
\begin{equation}
\label{g1_z0_z0}
\mathcal{G}({\bf x}, z_0,{\bf x}', z_0;p) =
\frac{6}{a^2} \frac{1}{(2\pi)^2} \int
\frac{ d{\bf q} \; e^{-i{\bf q}({\bf x}-{\bf x}')}
\sinh \left(z_0\sqrt{q^2 + 6p/a^2}\right)}{\sqrt{q^2 + 6p/a^2}
e^{z_0\sqrt{q^2 + 6p/a^2}}-k_0\sinh\left(z_0\sqrt{q^2 + 6p/a^2}\right)}
\end{equation}
\end{widetext}

\section{\label{app:all} Calculation of the higher order term in Eq. \ref{trp_all_final}}
Using the matrix notation of the Appendix
\ref{app:sequence_average}, we expand  $K^{-1}$ as
\begin{equation}
K^{-1} = \sum_{k=0}^{\infty} (\widehat{g} c)^k
\end{equation}

and insert it into (\ref{Z_repl_all}). Then, the additional term
from this expansion corresponding to the order $\Delta_p^2
\Delta_s^2$ will have the form:
\begin{equation} \exp \bigg[ \sigma_0^2 \xi_0^2 \int_0^N \!\! dn_1 \: dn_2 \: dn_3 \:
\widehat{g}(n_1,n_2) \: c(n_3-n_2) \: \beta \sum_{\alpha=1}^{m}
V(z_{\alpha}(n_3)) \bigg]
\end{equation}
This is equivalent to the following contribution to the
$\beta\Delta F/N$ before taking the replica limit $m \to 0$:
\begin{widetext}
\begin{equation}
\label{av3_same_repl}
\beta^3 \sigma_0^2 \xi_0^2 \int \! d{\bf x}' \: d{\bf x}'' \:
g({\bf x}'-{\bf x}'') \: \int_0^N \!\! dn_1 \: dn_2 \: dn_3 \:
c(n_3-n_2) 
\sum_{\alpha, \gamma, \eta =1}^{m}
\overline{V(z_{\alpha}(n_1))\delta({\bf x}_\alpha(n_1)-{\bf x}')
V(z_{\gamma}(n_2)) \delta({\bf x}_\gamma(n_2)-{\bf x}'')
V(z_{\eta}(n_3))}.
\end{equation}
\end{widetext}
Here, $\overline{\cdot}$ denotes the average in a reference system
of $m$ independent homogeneous replicated systems. 
First of all, we notice that only $m$ terms corresponding to
replicas with the same indices (i.e. $\alpha=\gamma=\eta$) will
contribute to $\Delta F$ in the transition point. After taking
the replica limit, we thus have
\begin{equation}
\label{av3_same}
\begin{split}
& \beta^3 \sigma_0^2 \xi_0^2 \int \! d{\bf x}' \: d{\bf x}'' \:
g({\bf x}'-{\bf x}'') \: \int_0^N \!\! dn_1 \: dn_2 \: dn_3 \:
c(n_3-n_2) \times \\
& [\overline{V(z(n_1))\delta({\bf x}(n_1)-{\bf x}')
V(z(n_2)) \delta({\bf x}(n_2)-{\bf x}'')
V(z(n_3))}]_0
\end{split}
\end{equation}

\begin{figure}[htb]
\begin{center}
\includegraphics[scale=0.4]{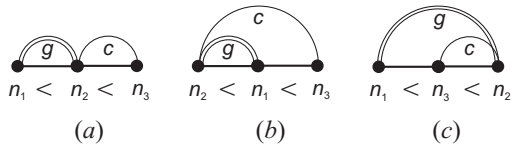}
\end{center}
\caption{ Diagrams contributing to the expression (\ref{av3_same}).
See text for explanation.
\label{diagrams}
}
\end{figure}

Note that the monomers $n_1$ and $n_2$ are "coupled" in the
expression (\ref{av3_same}) via the site-site correlation function
$g$, whereas the correlation between monomers $n_2$ and $n_3$ is
accounted for by the monomer-monomer correlation function
$c(n_3-n_2)$. Therefore, one has to take into account the
\emph{order} in the sequence of monomer numbers $n_1$, $n_2$, and
$n_3$. There are 3!=6 possible permutations of $n_1$, $n_2$, and
$n_3$ but in fact it is sufficient to consider the three possible
arrangements schematically shown in Fig.~\ref{diagrams}.
The other three are inverse of these ones. We
calculate the average in Eq.~(\ref{av3_same}) using the Green's
function of the reference system, like we did in Eq.~(\ref{V2_av_same}),
and then integrate it over $\int \! d{\bf x}'
\: d{\bf x}''$ and $\int_0^N \!\! dn_1 \: dn_2 \: dn_3$ . These
calculations are straightforward and give the following results
(in the dimensionless variables $\widetilde{\Gamma_s}$ and
$\widetilde{\Delta_s}$):
\begin{eqnarray}
(a) & = &  \Delta_p^2 \widetilde{\Delta_s}^2 \frac{36}{\sqrt{6 \Gamma_p}} \: \frac{1}{\widetilde{\Gamma_s}} \\
(b) & = &  \Delta_p^2 \widetilde{\Delta_s}^2 \frac{36}{\sqrt{6 \Gamma_p}} \: \frac{1}{\widetilde{\Gamma_s} + \sqrt{6 \Gamma_p}} \\
(c) & = &  \Delta_p^2 \widetilde{\Delta_s}^2 \frac{36}{\pi
\widetilde{\Gamma_s}^2} \: G^{22}_{22} \Big(
\frac{\widetilde{\Gamma_s}^2}{6 \Gamma_p} \Big|
\begin{array}{cc}1&1\\1/2&3/2 \end{array} \Big)
\end{eqnarray}
where $G$ is the Meijer function. Taking into account the double
counting (due to inverse sequences) and the prefactor
$\widetilde{\sigma_0}^2 \xi_0^2$ we arrive at the second term in
the expression (\ref{C_coeff}) for the coefficient $C$ in Eq.
(\ref{trp_all_final}). The first term in $C$ comes, as it has been
already noticed, from the expansion (\ref{exp_ln_det}).

%
%
%


\begin{thebibliography}{99}
\bibitem{footnote}
This is already obvious from the fact that for heteropolymers, the terms corresponding
to different replicas in $\Delta F$ do not vanish.
The contributions to $\Delta F$ which stem from only one replica can
be recovered from Eq.~(\ref{df_hetero_final}) in the limit $\Delta_s= \xi_0 $,
$ \Gamma_s \to 0$, after recognizing that
$\lim_{\Gamma_s \to 0} \Gamma_s q / (q^2 + \Gamma_s^2)^{3/2}$
has the properties of a delta-function on the semiaxis $q \in [0,\infty]$.
\end{thebibliography}
\end{document}